\documentclass[aps,prapplied, reprint, superscriptaddress,longbibliography]{revtex4-2}

\usepackage{amsmath}   
\usepackage{graphicx}  
\usepackage{hyperref}  
\usepackage{lineno}    

\usepackage{lineno,hyperref}
\usepackage{eurosym}
\usepackage{amsmath,amssymb,graphicx}
\usepackage{color}

\usepackage{subcaption}
\usepackage[justification=justified]{caption}
\usepackage{ragged2e}
\usepackage{booktabs}
\usepackage{array}    
\usepackage{float}
\newcolumntype{L}[1]{>{\raggedright\arraybackslash}p{#1}} 

\modulolinenumbers[5]
\newcommand{\bitem}{\begin{itemize}}
\newcommand{\fitem}{\end{itemize}}
\newcommand{\beq}{\begin{equation}}
\newcommand{\eeq}{\end{equation}}
\newcommand{\beqa}{\begin{eqnarray}}
\newcommand{\eeqa}{\end{eqnarray}}

\newcommand{\ket}[1]{|#1\rangle}                                
\newcommand{\bra}[1]{\langle#1|}                                

\newcommand{\UFSCar}{Departamento de Física, Universidade Federal de São Carlos, Rodovia Washington Luís, km 235 - SP-310, 13565-905 São Carlos, SP, Brazil}

\usepackage{orcidlink}

\usepackage{tikz}
\begin{document}

\title{Engineered Kerr Nonlinearities for Precise Quantum Control of Fock States}

\author{Gabriella G. Damas~\orcidlink{0000-0003-3376-9281}}
\email{gabriella.damasg@gmail.com}
\affiliation{Instituto de Física, Universidade Federal de Goiás, 74.001-970, Goiânia
- GO, Brazil}
\author{Ciro Micheletti~Diniz~\orcidlink{0000-0002-7602-0468}}
\affiliation{\UFSCar}

\author{Norton G. de Almeida~\orcidlink{0000-0001-8517-6774}}
\affiliation{Instituto de Física, Universidade Federal de Goiás, 74.001-970, Goiânia
- GO, Brazil}\affiliation{Instituto de Física de São Carlos, Universidade de São Paulo, Caixa Portal 369, 13560-970, São Carlos - SP, Brazil.}

\author{Celso J. Villas-Bôas~\orcidlink{0000-0001-5622-786X}}
\affiliation{\UFSCar}

\author{G. D. de Moraes Neto~\orcidlink{0000-0003-4273-8380}}
 \email{gdmneto@gmail.com}
\affiliation{Faculty of Civil Engineering and Mechanics, Kunming University of Science and Technology, Kunming, 650500, China}

\date{\today}

\begin{abstract}
We present a practical design framework for high-fidelity quantum control in coupled Kerr-nonlinear oscillators, directly addressing the challenge of spectral crowding. We show that systematic spectral degeneracies, which hinder selective addressing, are a direct consequence of rational Kerr-nonlinearity ratios ($K_1/K_2$). Our solution is a universal architectural principle: engineer this ratio to be a complex rational value, approximating an incommensurate number to systematically eliminate parasitic resonances. Using a Magnus expansion, we derive a complete effective Hamiltonian, including all Stark-shift corrections, to accurately target transitions. We numerically validate this framework by demonstrating protocols for the deterministic synthesis of NOON states, and high-photon-number Fock states, achieving ideal fidelities exceeding $\mathcal{F}>99.9\%$.  The protocols are shown to be robust against environmental decay and thermal effects. This work provides an architectural blueprint for bosonic processors in circuit QED and establishes foundational principles that could inform future designs of multi-mode quantum systems.
\end{abstract}

\maketitle

\section{INTRODUCTION}

The precise control of interacting quantum systems represents a central challenge in modern quantum science, with critical applications in quantum computation, simulation, and metrology \cite{Krantz2019,Preskill2018,chitambar2019quantum,streltsov2017colloquium,acin2018quantum,nielsen2010quantum,Giovannetti2006,Giovannetti2011,degen2017quantum,QKD2009Review,gisin2007quantum,Simulation2014Review}. Networks of coupled bosonic modes have emerged as particularly promising platforms, experimentally realized in superconducting circuits \cite{Wallraff2004,Kjaergaard2020,Blais2021,Kudra2025}, trapped ions \cite{Home2009,Bruzewicz2019}, and optomechanical systems \cite{Aspelmeyer2014}. The infinite-dimensional Hilbert space of bosonic modes provides a powerful resource for hardware-efficient quantum error correction through bosonic codes \cite{Gottesman2001,Michael2016,Ofek2016,Lescanne2020}, enabling demonstrations of fault-tolerant quantum operations ~\cite{Ofek2016,Hu2019,Fluhmann2019}.

A fundamental requirement for harnessing these capabilities is the ability to \emph{selectively address individual quantum transitions}. In single-mode light-matter systems governed by the Jaynes-Cummings model~\cite{prado2013engineering,rosado2015upper}, the quantum Rabi model~\cite{cong2020}, and the Dicke model~\cite{mu2020dicke}, such selectivity is well established. Through the strategic application of Stark-shifting fields, the resonance conditions for Rabi oscillations can be precisely tailored to specific Jaynes-Cummings doublets while keeping all other transitions in a dispersive regime. This principle underpins protocols for state preparation and tomography in cavity quantum electrodynamics (QED)~\cite{Franca01,VillasBoas2019, Vrajitoarea2020} and trapped ions~\cite{Solano00,Solano05,Franca05,Rossetti2014}.

Extending this selective control to multimode systems presents substantially greater challenges. As the number of modes increases, the growing density of transition frequencies leads to \emph{spectral crowding}, where control fields inevitably excite off-resonant transitions. Recent mitigation efforts have focused on sophisticated control techniques at the pulse and ancilla level. As reviewed in Ref.~\cite{ma2021quantum}, a common strategy involves coupling the bosonic modes to a transmon ancilla and using it as a nonlinear resource. Quantum optimal control methods, for instance, can design pulses that are robust against ancilla-mediated crosstalk, where the state of a spectator mode degrades operations on a target mode \cite{you2024crosstalk}. Similarly, driven ancillas can be used to engineer arbitrary photon-number-dependent Hamiltonians for tasks such as Kerr cancellation and noise-resilient gates \cite{wang2021photonnumberdependent, Heeres2015}.

While these methods provide valuable control-layer tools, they are designed to work around a given system Hamiltonian and do not address a more fundamental limitation: systematic spectral degeneracies inherent to the coupled oscillator dynamics~\cite{Yusipov2023}. These intrinsic degeneracies create unavoidable resonances for parasitic transitions, a problem fundamentally insoluble through pulse shaping alone. To our knowledge, a systematic framework for achieving high-fidelity selective control between coupled bosonic modes through architectural design has remained an unmet critical need.

In this work, we address this gap by developing a comprehensive framework for \emph{selective quantum control in coupled Kerr-nonlinear resonators}, grounded in the Floquet-Magnus expansion. 
Our approach yields a time-independent effective Hamiltonian that systematically incorporates higher-order coherent errors, providing closed-form analytic expressions for Stark shifts and effective drive strengths across all parameter regimes. A central insight emerging from our analysis is a universal design principle for spectral crowding mitigation: engineering an \emph{irrational ratio of Kerr nonlinearities} between coupled modes. This strategy systematically eliminates accidental degeneracies that obstruct selective addressability, providing a scalable approach for bosonic architectures \cite{Yusipov2023}.

We validate this framework through numerical simulations of beam-splitter and two-mode squeezing gates implemented in Kerr-nonlinear resonators. Using experimentally accessible parameters from state-of-the-art circuit QED devices~\cite{Ye2020,Wang2016,Grimm2020,Ma2021}, we demonstrate that the proposed design principle enables high-fidelity selective transitions while suppressing off-resonant leakage and Stark-induced phase errors. Beyond coherent control, the same principles can be extended to dissipative stabilization of entangled states via engineered reservoirs~\cite{deMoraesNeto2014,Prado2014}, highlighting the generality of the approach. The Floquet–Magnus expansion further provides a systematic means to identify and compensate coherent error sources beyond the rotating-wave approximation, establishing a direct connection between the system’s nonlinear architecture and its controllability.

This article is structured as follows: Section \ref{sec:theory} presents our theoretical framework for selective control in coupled Kerr-nonlinear oscillators. Section \ref{sec: app engineering} demonstrates quantum state engineering applications, including NOON states and high-photon-number Fock states. In Section \ref{sec:experimental}, we discuss experimental considerations and fidelity analysis in superconducting circuit platforms, validating the practical feasibility of our framework. Appendices provide technical details of error analysis and numerical methods. 

\section{THEORETICAL FRAMEWORK FOR SELECTIVE CONTROL}
\label{sec:theory}

This section establishes the theoretical basis for selective control in coupled Kerr-nonlinear oscillators. Starting from the fundamental Hamiltonian, we systematically analyze the transition spectrum, derive the conditions for high-fidelity operations, and develop a comprehensive framework for overcoming spectral crowding through careful parameter engineering.




\begin{figure*}[t]
\centering
\includegraphics[width=\linewidth]{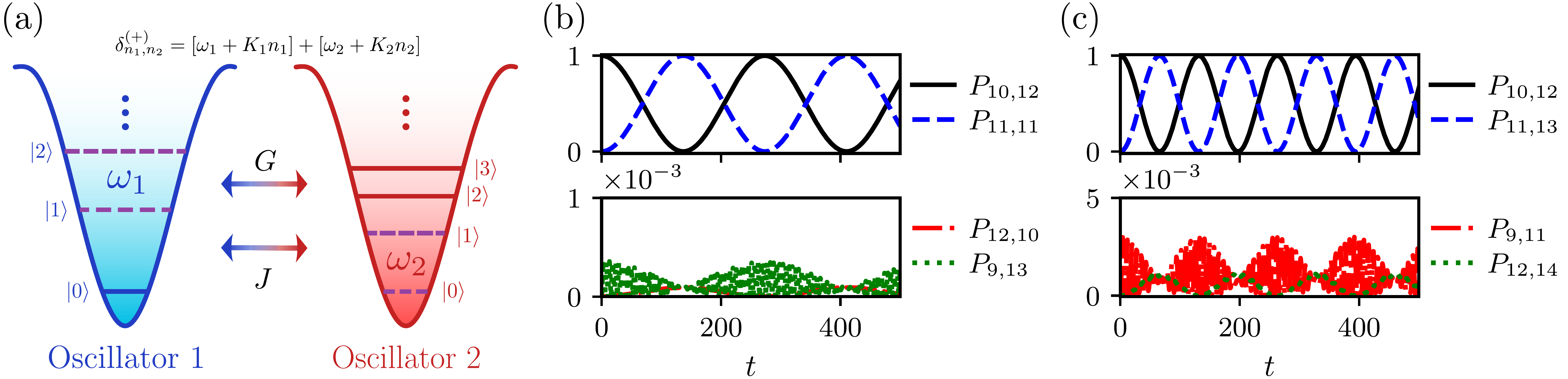}
\caption{\justifying Selective quantum control of individual fock-state transitions. (a) Pictographic representation of the model described by Eq.~\eqref{eq:H_full}. In this example, the resonance is between the state $\ket{1,0}$ and $\ket{2,1}$ (dashed purple lines), with detuning transition $\delta^{(+)}_{n_1,n_2}$ given by Eq.~\eqref{eq:delta_plus}. (b) Population dynamics under a resonant drive on the BS transition $\ket{10, 12} \leftrightarrow \ket{11, 11}$. The upper plot shows Rabi oscillations between the target states, while the lower plot shows the leakage population into the off-resonant neighboring states $\ket{9, 13}$ and $\ket{12, 10}$. (c) Dynamics under a resonant drive on the TMS transition $\ket{10, 12} \leftrightarrow \ket{11, 13}$. The lower plot shows the leakage into the neighboring states $\ket{9, 11}$ and $\ket{12, 14}$. In both cases, the strong suppression of off-resonant transitions demonstrates the high fidelity of state-selective addressing.}
\label{fig:trio}
\end{figure*}


\subsection{System Hamiltonian and Spectral Structure}

We consider a system of two coupled Kerr-nonlinear oscillators, Fig.~\ref{fig:trio} (a), a paradigmatic model that describes numerous quantum platforms, including superconducting circuits, trapped ions, and optomechanical systems. The system's dynamics is governed by the Hamiltonian $\hat{H}_{\mathrm{Kerr}}$, expressed in natural units ($\hbar=1$) as:
\begin{equation}
\hat{H}_{\mathrm{Kerr}} = \hat{H}_0 + \hat{H}_I,
\label{eq:H_full}
\end{equation}
where the free Hamiltonian $\hat{H}_0$ and interaction Hamiltonian $\hat{H}_I$ are given by:
\begin{align}
\hat{H}_0 &= \sum_{j=1}^2\left[\omega_j\hat{n}_j + \frac{K_j}{2}\hat{n}_j(\hat{n}_j-1)\right], \\
\hat{H}_I &= J(\hat{a}_1^\dagger\hat{a}_2 + \hat{a}_1\hat{a}_2^\dagger) + G(\hat{a}_1^\dagger\hat{a}_2^\dagger + \hat{a}_1\hat{a}_2).
\end{align}
For the $j$-th mode ($j = 1, 2$), $\hat{a}_j$ ($\hat{a}_j^\dagger$) denotes the bosonic annihilation (creation) operator, $\hat{n}_j = \hat{a}_j^\dagger \hat{a}_j$ is the number operator, $\omega_j$ represents the fundamental frequency, and $K_j$ is the self-Kerr coefficient. The interaction Hamiltonian contains two fundamental processes: a beam-splitter (BS) interaction of strength $J$ that conserves total photon number, and a two-mode squeezing (TMS) interaction of strength $G$ that creates or destroys photon pairs.  On the other hand, in the free Hamiltonian, the Kerr term $\frac{K_j}{2}\hat{n}_j(\hat{n}_j-1)$ is mathematically equivalent to the normally ordered form $\frac{K_j}{2}\hat{a}_j^{\dagger 2}\hat{a}_j^2$, up to a constant frequency shift of $K_j/2$ in $\omega_j$, both yielding identical eigenvalues $K_j n_j(n_j-1)/2$ in the Fock basis. We employ the $\hat{n}_j(\hat{n}_j-1)$ form for its transparent diagonal structure.

In circuit QED implementations, which serve as our primary experimental motivation, the Kerr nonlinearity $K_j$ arises from the intrinsic anharmonicity of transmon qubits due to the cosine potential of Josephson junctions~\cite{Koch2007}. The BS interaction can be realized through capacitive coupling or tunable couplers~\cite{Yan2018}, while the TMS interaction is typically activated by parametric driving of coupling elements at sum-frequency conditions~\cite{Puri2017}.

To analyze the spectral structure and identify conditions for selective control, we transform to the interaction picture with respect to $\hat{H}_0$. The resulting time-dependent interaction Hamiltonian, expressed in the two-mode Fock basis $\{\ket{n_1, n_2}\}$, reveals the fundamental transition structure:
\begin{align} \tilde{H}_I(t) = &\sum_{n_1,n_2} \Omega^{(-)}_{n_1,n_2} \times \nonumber \\ &\qquad \left( e^{i\delta^{(-)}_{n_1,n_2} t} |n_1+1, n_2-1\rangle \langle n_1, n_2| + \mathrm{H.c.} \right) \nonumber \\ + &\sum_{n_1,n_2} \Omega^{(+)}_{n_1,n_2} \times \nonumber \\ &\qquad \left( e^{i\delta^{(+)}_{n_1,n_2} t} |n_1+1, n_2+1\rangle \langle n_1, n_2| + \mathrm{H.c.} \right). \label{eq:H_int_fock_expanded} \end{align}
where the detunings depend linearly on the Kerr nonlinearities,
\begin{align}
\delta^{(-)}_{n_1,n_2} &= [\omega_1 + K_1 n_1] - [\omega_2 + K_2 (n_2-1)], \\
\delta^{(+)}_{n_1,n_2} &= [\omega_1 + K_1 n_1] + [\omega_2 + K_2 n_2]. \label{eq:delta_plus}
\end{align}
Each term in Eq.~\eqref{eq:H_int_fock_expanded} corresponds to a specific quantum transition characterized by the number-dependent Rabi frequencies,
\begin{align}
    \Omega^{(-)}_{n_1,n_2} &= J\sqrt{(n_1+1)n_2}, \\
    \Omega^{(+)}_{n_1,n_2} &= G\sqrt{(n_1+1)(n_2+1)}.
\end{align}

This spectral structure forms the foundation for selective control. This number-dependency, however, is a double-edged sword: while it provides the potential for unique spectral addresses, it also creates the fundamental challenge of spectral crowding. Even with only two modes, the infinite Fock-space ladders contain an infinite number of potential transitions. If system parameters (e.g., $K_1, K_2$) are not chosen carefully, distinct transitions, even those far apart in photon number, can possess nearly identical frequencies ($\delta \approx \delta'$). This accidental degeneracy is the core problem that must be solved.

To achieve the ultra-high fidelities required for quantum computation, one must therefore not only drive resonantly but also rigorously account for coherent errors arising from these off-resonant (or accidentally near-resonant) terms. We accomplish this by deriving a time-independent effective Hamiltonian via the Floquet-Magnus expansion (see Appendix \ref{appendix: A} for the full derivation, which systematically quantifies effects such as AC Stark shifts and state leakage). To validate this approach, we perform numerical simulations of the full system dynamics under $\hat{H}_{\mathrm{Kerr}}$ to demonstrate that high-fidelity selective control is achievable with a careful choice of parameters. Figure~\ref{fig:trio} shows two such examples for high photon numbers, one for a BS transition (b) and one for a TMS transition (c). In both cases, the population remains confined to the target states, with leakage to neighboring states suppressed below $0.5\%$, confirming that precise spectral addressing is within reach.

\subsection{Engineering Selectivity: Avoiding Accidental Degeneracies}

A fundamental challenge in multi-mode quantum control is spectral crowding, the phenomenon where multiple distinct quantum transitions become accidentally degenerate, causing catastrophic failure of selectivity when a drive simultaneously excites multiple transitions. As introduced above, this problem persists even in two-mode systems due to the infinite size of the Fock space.

The conditions for such degeneracies emerge directly from the detuning structure. For a BS drive resonant with the $\ket{n_0, m_0} \leftrightarrow \ket{n_0+1, m_0-1}$ transition, the relative detuning for any parasitic BS transition $\ket{n', m'} \leftrightarrow \ket{n'+1, m'-1}$ is:
\begin{equation}
\delta'_{\text{rel}} = K_1(n' - n_0) - K_2(m' - m_0).
\label{eq: rel_kerr}
\end{equation}
An accidental degeneracy occurs when $\delta'_{\text{rel}} \approx 0$ for $(n', m') \neq (n_0, m_0)$, which implies:
\begin{equation}
\frac{K_1}{K_2} \approx \frac{m' - m_0}{n' - n_0} = \frac{\Delta m}{\Delta n}.
\label{eq:degeneracy_condition}
\end{equation}

The critical insight from Eq.~\eqref{eq:degeneracy_condition} is that if $K_1/K_2$ is a rational number $p/q$, then systematic degeneracies occur for the infinite family of transitions where $(\Delta n, \Delta m) = (kq, kp)$ for any integer $k$. This creates a dense network of unresolvable transitions throughout the Fock space.

\begin{figure*}[t]
\centering
\includegraphics[width=\linewidth]{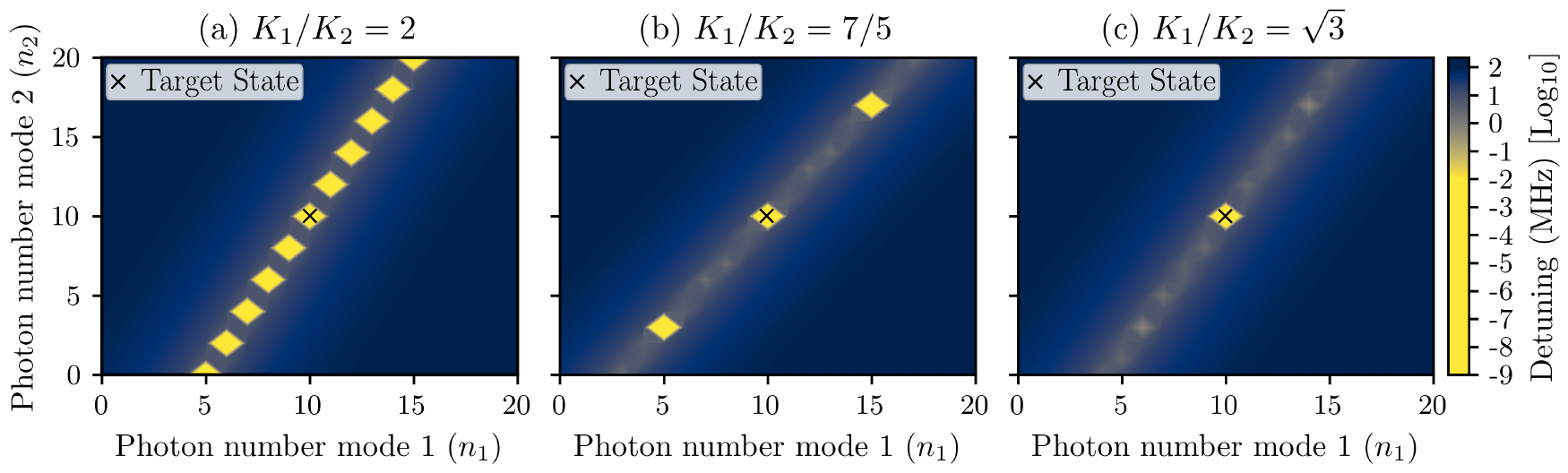}
\caption{\justifying Impact of kerr coefficient ratio on spectral addressability. The plots illustrate the detuning of parasitic transitions relative to a target state (yellow square) within a two-mode Fock space, defined by photon numbers in mode 1 ($n_1$) and mode 2 ($n_2$). The color indicates the detuning of each state's transition frequency from the target, with darker regions signifying near-degeneracy or spectral crowding. (a) With a integer ratio Kerr ratio, $K_1/K_2 = 2$, systematic degeneracies emerge, creating dense clusters of near-resonant transitions that hinder unique addressability. (b) For a rational ratio, $K_1/K_2 = 7/5$, the degeneracy pattern is more intricate but remains highly structured. (c) When the ratio is incommensurate, $K_1/K_2 = \sqrt{3}$, systematic degeneracies are eliminated. This results in a uniform distribution of transition frequencies, allowing the target state to be addressed with high fidelity and minimal off-resonant excitation of neighboring states.}

\label{fig:spectral_crowding}
\end{figure*}

This analysis yields a crucial design principle: to eliminate systematic degeneracies and ensure unique spectral addresses, the ratio of Kerr coefficients $K_1/K_2$ should ideally be an incommensurate number. With incommensurate Kerr coefficients, no exact degeneracies exist between distinct transitions, though near-degeneracies may still occur and must be analyzed case by case. 

However, any experimental implementation will necessarily be limited to a rational approximation of this ideal. This leads us to the practical design principle: one must engineer a rational ratio $K_{1}/K_{2} = p/q$ where $p$ and $q$ are coprime and sufficiently large integers. By doing so, the first family of systematic degeneracies (as per Eq. 10, occurring at $\Delta n = q, \Delta m = p$) is pushed far from the low-photon-number regime, effectively creating a large, degeneracy-free manifold for computation.

The visual evidence in Fig.~\ref{fig:spectral_crowding} strongly supports this principle, showing how rational ratios create systematic patterns of spectral crowding while incommensurate ratios break these systematic patterns, resulting in a far more sparse distribution of resonances.

The overarching condition for high-fidelity selectivity requires that the detuning of any parasitic transition substantially exceeds its coupling strength:
\begin{equation}
|\delta'{\text{rel}}| \gg \Omega^{(-)}_{n',m'} = J\sqrt{(n'+1)m'}.
\label{eq:selectivity_condition}
\end{equation}
This ensures that when a specific transition is brought into resonance, all other transitions remain strongly suppressed. The uniform distribution achieved with incommensurate Kerr ratios makes this condition easier to satisfy across the entire Fock space, as it prevents the formation of dense clusters of near-degenerate transitions that would otherwise require impractically narrow bandwidths for selective addressing.

The efficacy of incommensurate ratios depends on their approximability by rationals. Poorly approximable ratios (e.g., the golden ratio $\phi$) provide superior performance by maximizing the minimum detuning $\delta'_{\text{rel}}$ in Eq. (\ref{eq: rel_kerr}) across the Fock space. In contrast, simple rational ratios create systematic degeneracies, degrading fidelity through resonant leakage. This analysis provides a theoretical foundation for selecting ratios that maximize spectral addressability in practical implementations.

\subsection{Generalized Framework: Incorporating Cross-Kerr Coupling}

A more comprehensive physical model for coupled anharmonic oscillators includes the cross-Kerr interaction $\hat{H}_{CK} = K_{12}\hat{n}_1\hat{n}_2$, which is added to the free Hamiltonian $\hat{H}_0$. This term describes the mutual frequency shift between modes dependent on their respective photon numbers. This term naturally arises in circuit QED from Josephson junction nonlinearities and becomes particularly significant in strongly coupled regimes \cite{Vrajitoarea2020, you2024crosstalk}. 

The cross-Kerr interaction modifies the eigenenergy structure and consequently alters the detuning conditions for quantum transitions. For a BS transition, the generalized relative detuning, including cross-Kerr effects, becomes:
\begin{equation}
\delta'{\text{rel}} = K_1 \Delta n - K_2 \Delta m + K_{12}(\Delta m - \Delta n),
\label{eq:detuning_cross_kerr}
\end{equation}
where $\Delta n = n' - n_0$ and $\Delta m = m' - m_0$ represent the differences in photon numbers from the target state.

The degeneracy condition $\delta'{\text{rel}} \approx 0$ now involves three parameters:
\begin{equation}
K_1 \Delta n - K_2 \Delta m + K_{12}(\Delta m - \Delta n) \approx 0.
\end{equation}

This leads to a more sophisticated design principle for robust quantum control: to eliminate all systematic degeneracies, the set of Kerr coefficients ${K_1, K_2, K_{12}}$ must be linearly independent over the rational numbers. That is, there should exist no nonzero integers ${c_1, c_2, c_3}$ satisfying:
\begin{equation}
c_1 K_1 + c_2 K_2 + c_3 K_{12} = 0.
\end{equation}

The inclusion of cross-Kerr coupling presents both challenges and opportunities for quantum control. While it introduces additional complexity to the spectral structure and potential new pathways for spectral crowding, it also provides an extra engineering degree of freedom that can be exploited to lift degeneracies that would otherwise be unavoidable with only self-Kerr nonlinearities. In practical implementations, careful characterization and compensation of cross-Kerr effects are essential for achieving the highest fidelity operations.

This comprehensive theoretical framework establishes the fundamental principles for selective control in coupled Kerr-nonlinear systems, providing the foundation for the experimental implementations discussed in Section~\ref{sec:experimental}.


\section{Applications in Quantum State Engineering}
\label{sec: app engineering}

Quantum state engineering leverages tailored interactions to synthesize nonclassical states for quantum technologies~\cite{Haroche2006}. The framework presented in this work provides a platform for such control, enabling deterministic Fock-state manipulation in both closed and open quantum systems. The versatility of this approach allows for a wide range of protocols, from simple state transfers to the dissipative stabilization of bosonic codes, with key examples discussed below and summarized in Table~\ref{tab: preparation protocols}. These applications have direct relevance in metrology, error correction, and simulation~\cite{Krantz2019}.

\begin{table*}[bth!]
\centering
\caption{Summary of State Preparation Protocols. (Typical parameters: $J/2\pi \approx 10$ MHz, $\kappa_2/2\pi \approx 1$ MHz, $\kappa_1/2\pi \approx 10$ kHz)}
\label{tab: preparation protocols}
\begin{tabular}{@{} l c c l l @{}}
\toprule
\textbf{Protocol} & \textbf{Initial State} & \textbf{Target State} & \textbf{Operation(s)} & \textbf{Ancilla Req.} \\
\midrule
Photon Transfer & $\ket{n_0, m_0}$ & $\ket{n_0+1, m_0-1}$ & BS $\pi$-pulse & Initialized \\
Superposition & $\ket{n_0, m_0}$ & $c_1\ket{n_0,m_0}+c_2\ket{n_0+1,m_0-1}$ & BS partial pulse & Initialized \\
Entangled Pair & $\ket{n_0, m_0}$ & $\ket{n_0+1, m_0+1}$ & TMS $\pi$-pulse & - \\
Entangled Superposition & $\ket{0, 0}$ & $\frac{1}{\sqrt{2}}(\ket{0,0}-i\ket{1,1})$ & TMS $\pi/2$-pulse & - \\
Two-Photon NOON & $\ket{0,0}$ & $\frac{1}{\sqrt{2}}(\ket{2,0} + e^{i\phi}\ket{0,2})$ & 4-pulse seq. & - \\
Fock State ($N=4$) & $\ket{0,0}$ & $\ket{4,0}$ & Double TMS-BS & - \\
Fock Stabilization & $\ket{0,0}$ & $\ket{n_0,0}$ & Dissipative Pumping & Strongly damped \\
\bottomrule
\end{tabular}
\end{table*}

\subsection{Unitary State Preparation in Closed Systems}

In the closed-system limit, the effective Hamiltonians generate coherent, unitary dynamics within a well-defined two-level subspace. The interaction time $t$ serves as a control parameter for rotations on the Bloch sphere of the effective qubit.

\subsubsection{Applications of the Selective Beam-Splitter}

The Hamiltonian $H_{\mathrm{BS}}^{\mathrm{eff}} = \Omega_{\mathrm{BS}} ( |n_0+1, m_0-1 \rangle \langle n_0, m_0 | + \text{H.c.} )$ governs transitions within the number-conserving subspace $\{\ket{n_0,m_0}, \ket{n_0+1,m_0-1}\}$.

\begin{itemize}
    \item \textbf{Deterministic Photon Transfer:} Complete control over single-photon number states is realized via resonant $\pi$-pulses. For example, the transformation $\ket{n_0,m_0} \to \ket{n_0+1,m_0-1}$ is achieved by evolving for a time $t=\pi/\Omega_{\mathrm{BS}}$, where the Rabi frequency is $\Omega_{\mathrm{BS}}=J\sqrt{(n_0+1)m_0}$.
    
    \item \textbf{Generation of Fock-State Superpositions:} Identifying the states $\{\ket{n_0}, \ket{n_0+1}\}$ with logical qubit states $\{\ket{0}_L,\ket{1}_L\}$, arbitrary single-qubit rotations can be implemented using an ancilla mode~\cite{Law1996}. An rotating gate $R_x(\theta)$ gate corresponds to an interaction time $t=\theta/(2\Omega_{\mathrm{BS}})$. In particular, a $\pi/2$-pulse prepares an equal superposition, such as transforming $\ket{n_0,1}$ into the state $\frac{1}{\sqrt{2}} (\ket{n_0,1} - i\ket{n_0+1,0})$.
    
    \item \textbf{Synthesis of Arbitrary Superpositions:} Controlled sequences of partial SWAP operations allow the construction of arbitrary superpositions of Fock states within a number-conserving manifold. For instance, the (normalized) binomial state $\frac{1}{2}(\sqrt{3}\ket{1,2}+\ket{3,0})$ can be synthesized from an initial state $\ket{0,3}$ via a sequence of appropriately timed swaps.
\end{itemize}

\subsubsection{Applications of Selective Two-Mode Squeezing}

The interaction $H_{\mathrm{TMS}}^{\mathrm{eff}} = \Omega_{\mathrm{TMS}} ( | n_0+1, m_0+1\rangle \langle n_0, m_0| + \text{H.c.} )$ selectively couples $\ket{n_0,m_0}$ and $\ket{n_0+1,m_0+1}$, allowing the creation and annihilation of controlled pairs.   

\begin{itemize}
    \item \textbf{Deterministic Generation of Two-Mode Fock States:} Starting from the vacuum state $\ket{0,0}$, a resonant $\pi$-pulse on the $|0,0\rangle \leftrightarrow |1,1\rangle$ transition deterministically prepares the product state $\ket{1,1}$. By iteratively applying this process and tuning the drive frequency at each step, one can deterministically climb the two-mode Fock ladder to prepare any state of the form $\ket{n,n}$.
    
    \item \textbf{Generation of Entangled Superpositions:} Partial pulses generate entangled superpositions. For example, a $\pi/2$-pulse applied to $\ket{0,0}$ yields the state $\frac{1}{\sqrt{2}} (\ket{0,0} - i\ket{1,1})$. This protocol can be extended to prepare states of the form $\frac{1}{\sqrt{2}}(\ket{0,0}+e^{i\phi}\ket{n,n})$, which are a valuable resource for quantum metrology aiming to surpass the standard quantum limit~\cite{Giovannetti2004}.
\end{itemize}

In summary, the selective interactions provide a complete set of tools for deterministic, unitary control over the quantum state in the Fock basis.

\subsection{Autonomous State Preparation in Open Systems}

In the presence of dissipation, selective interactions can be exploited to engineer system–environment couplings such that the target state emerges as the unique steady state of the dynamics. This dissipative approach provides robustness against control imperfections and enables autonomous error correction via reservoir engineering~\cite{Poyatos1996}.

\subsubsection{Dissipative Fock-State Stabilization}

A pure Fock state $\ket{n_0}$ of a high-quality mode (mode~1, decay rate $\kappa_1$) can be stabilized via coupling to a strongly damped ancilla mode (mode~2, decay rate $\kappa_2$). The required rate hierarchy is $\kappa_2 \gg J \gg \kappa_1$, which ensures the ancilla acts as an effective reservoir. A classical drive on the ancilla, together with a selective BS interaction resonant with $|n_0-1,1\rangle \leftrightarrow |n_0,0\rangle$, pumps excitations into the target mode. Excess photons in mode~1 are removed by an additional selective interaction resonant with $|n_0+1,0\rangle \leftrightarrow |n_0,1\rangle$, which transfers them into the lossy ancilla where they rapidly decay. The combination of these engineered pumping and cooling channels yields $\ket{n_0}_1$ as the unique, dynamically-maintained steady state~\cite{Holland1996}.

\subsubsection{Stabilization of Entangled States}

Combining the selective TMS interaction with local dissipation provides a mechanism for stabilizing entangled steady states. The interplay between coherent pair creation driven by $H_{\mathrm{TMS}}^{\mathrm{eff}}$ and incoherent single-photon loss events can be engineered to produce a two-mode squeezed vacuum as the unique steady state. The inclusion of the Kerr nonlinearities is crucial for two reasons: they provide the spectral addressability to isolate the desired TMS interaction, and they prevent runaway parametric amplification that would otherwise occur in a linear system. This approach enables the stabilization of more exotic nonclassical states such as entangled Schrödinger cat states, e.g., $\frac{1}{\sqrt{2}}(\ket{\alpha,\alpha}+\ket{-\alpha,-\alpha})$~\cite{Puri2017}. The engineered dissipation continuously corrects for errors such as photon loss, thereby protecting the delicate quantum coherence of these otherwise fragile states.

Thus, by harnessing dissipation, these selective interactions enable the autonomous preparation and protection of valuable nonclassical states.

\subsection{Advanced Applications in Quantum Information and Simulation}

The effective BS and TMS Hamiltonians provide a set of primitive operations for a broad range of applications beyond basic state preparation.

\begin{itemize}
    \item \textbf{Quantum Walks in Fock Space:} Sequences of selective $\pi$-pulse BS operations can simulate the hopping of an excitation along a Fock-state lattice, enabling the study of transport phenomena and search algorithms in a synthetic dimension~\cite{Longhi2011}.

    \item \textbf{Heralded Non-Gaussian State Engineering:} Weak BS coupling between a target mode and a measured ancilla heralds photon subtraction or addition, providing a probabilistic route to non-Gaussian states from arbitrary initial states~\cite{Ourjoumtsev2006}.

    \item \textbf{Deterministic Synthesis of Bosonic Quantum Codes:} Precise sequences of selective interactions can deterministically prepare bosonic error-correcting code states, such as binomial, cat, or Gottesman-Kitaev-Preskill (GKP) codes, which are central to fault-tolerant quantum computing~\cite{Gottesman2001, Michael2016}.

    \item \textbf{Analog Quantum Simulation of Spin Models:} By mapping Fock-state subspaces to effective spin degrees of freedom, the BS and TMS interactions realize effective XX and XX+YY spin couplings, respectively, enabling the simulation of lattice spin models~\cite{Hartmann2006}.
\end{itemize}

These advanced applications highlight the role of selective interactions as fundamental building blocks for scalable quantum technologies.

\subsection{Expanded Protocol: Deterministic NOON-State Preparation}

\begin{figure*}[t] 
    \centering
    \includegraphics[width=\textwidth]{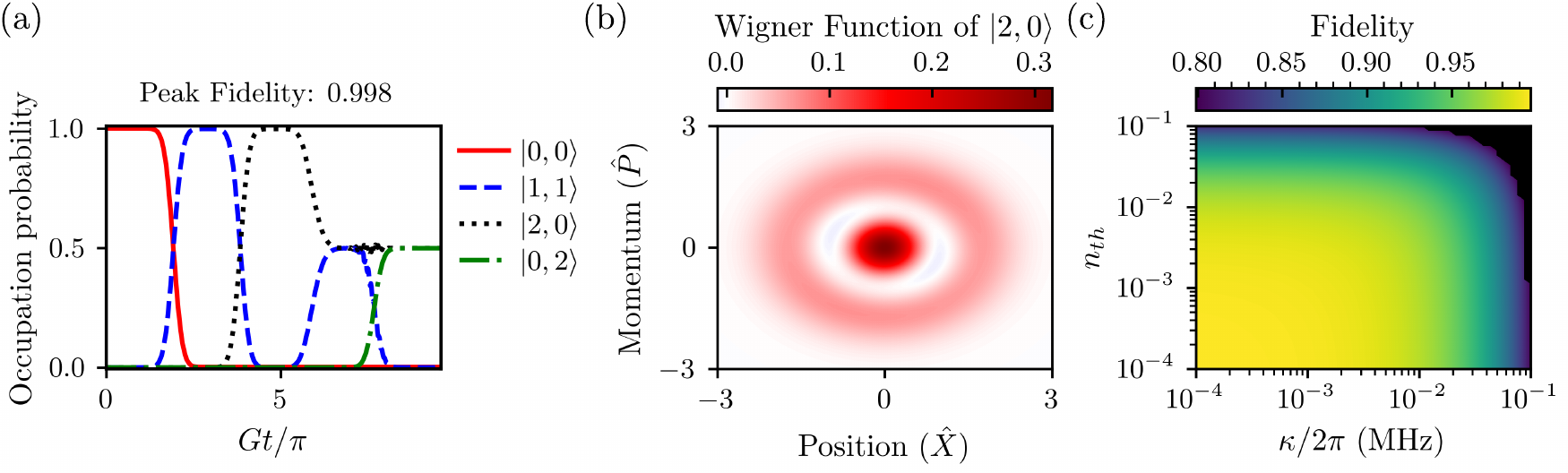}
    \caption{\justifying Direct synthesis and robustness of a two-photon NOON state protocol.
    (a) Population dynamics of the sequential operations, demonstrating the generation of the path-entangled state $(\ket{2,0} + \ket{0,2})/\sqrt{2}$ from the vacuum $\ket{0,0}$. A peak fidelity of 0.998 with the target state is achieved in the ideal case, i.e., without dissipation or temperature.
    (b) Wigner function of the first mode after tracing over the second mode of the NOON state $(\ket{2,0} + \ket{0,2})/\sqrt{2}$, showing the phase-space distribution of the reduced state $\rho_1 = \frac{1}{2}(\ket{0}\bra{0} + \ket{2}\bra{2})$.
    (c) Protocol robustness against decoherence, mapped as the peak fidelity versus decay rate $\kappa$ and thermal occupation $n_{th}$. The high-fidelity operating regime is highlighted; fidelities below 0.80 are shown in black.
    Key parameters: $G/2\pi = J/2\pi = 20$ MHz, $K_1/2\pi = 300$ MHz, and $K_2/2\pi \approx 212$ MHz.
}
    \label{fig:noon}
    \end{figure*}

An especially relevant application is the deterministic synthesis of NOON (N00N) states,
\begin{equation}
    \ket{\text{NOON}}_N = \frac{1}{\sqrt{2}} \left( \ket{N,0} + e^{i\phi}\ket{0,N} \right),
\end{equation}
which are a key resource for quantum-enhanced metrology, enabling phase measurements with sensitivity at the Heisenberg limit~\cite{Dowling2008}.

\paragraph*{Protocol.}
We demonstrate deterministic preparation of the two-photon NOON state $\frac{1}{\sqrt{2}}(\ket{2,0} + \ket{0,2})$ through the following pulse sequence:
\begin{enumerate}
    \item \textbf{Initial Entanglement:} A resonant $\pi/2$-pulse with the TMS interaction applied to $\ket{0,0}$ generates the entangled superposition $(\ket{0,0}-i\ket{1,1})/\sqrt{2}$.
    \item \textbf{Photon Consolidation:} A selective BS $\pi$-pulse, resonant with the $|1,1\rangle \leftrightarrow |2,0\rangle$ transition, transforms $\ket{1,1}$ to $\ket{2,0}$, yielding $(\ket{0,0}-i\ket{2,0})/\sqrt{2}$.
    \item \textbf{Ancilla Re-initialization:} The ancilla mode is reset to vacuum while preserving the target mode state, converting the system to $\frac{1}{\sqrt{2}}(\ket{0,0}-i\ket{2,0})$.
    \item \textbf{Mode Symmetrization:} A second TMS $\pi/2$-pulse generates entanglement between the $\ket{0,0}$ and $\ket{1,1}$ components, followed by a BS $\pi$-pulse on $|1,1\rangle \leftrightarrow |0,2\rangle$ to produce the final NOON state $\frac{1}{\sqrt{2}}(\ket{2,0} + e^{i\phi}\ket{0,2})$.
\end{enumerate}

The performance of this protocol is illustrated in Fig.~\ref{fig:noon}. Panel (a) shows the population dynamics under the optimized pulse sequence, confirming high-fidelity preparation of the two-photon NOON state with peak fidelity of 0.998. Panel (b) displays the Wigner function of the reduced state of the first mode, showing the characteristic phase-space distribution of the statistical mixture $\frac{1}{2}(\ket{0}\bra{0} + \ket{2}\bra{2})$ that results from tracing over one mode of the entangled state. Panel (c) quantifies the protocol's robustness to dissipation and thermal noise, demonstrating fidelities above 0.95 for decay rates up to $\kappa/2\pi\sim100$ kHz and thermal occupations $n_{th}<0.08$.

\paragraph*{Remarks.}
This construction provides a deterministic and scalable route to high-N path-entangled states, circumventing the probabilistic limitations of conventional linear-optical schemes. The ability to control the phase $\phi$ through the relative phase of the drive fields further enhances the metrological versatility of the generated state.

\subsection{Expanded Protocol: Deterministic Fock State Synthesis}

\begin{figure*}[t] 
    \centering
    \includegraphics[width=\textwidth]{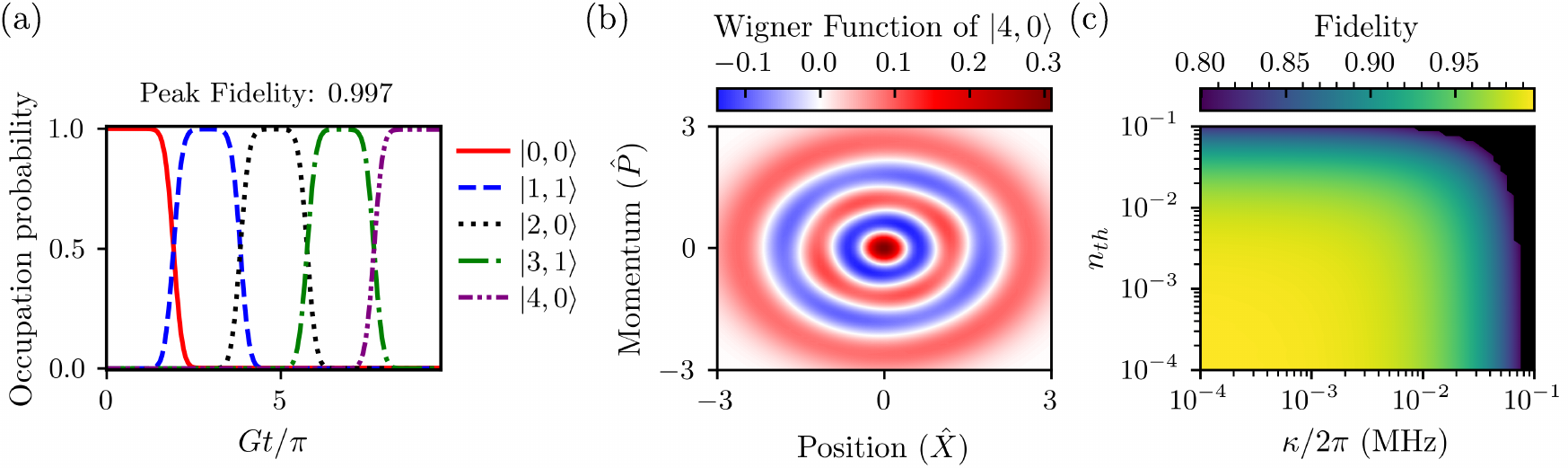}
 \caption{\justifying Synthesis and robustness of a $|4\rangle$ Fock state. (a) Population dynamics of the sequential protocol, demonstrating the state transfer from vacuum $|0,0\rangle$ to the target $|4,0\rangle$ with a peak fidelity of 0.997 in the ideal case. (b) Wigner function of the first mode, showing the characteristic phase-space distribution of the $|4\rangle$ Fock state. The central positive peak and concentric rings confirm the high-quality preparation of the state. (c) Protocol robustness against decoherence, mapped as the peak fidelity versus decay rate $\kappa$ and thermal occupation $n_{\text{th}}$. The high-fidelity operating regime is highlighted; fidelities below 0.8 are shown in black. Key parameters: $G/2\pi = J/2\pi = 20$ MHz, $K_1/2\pi = 300$ MHz, and $K_2/2\pi \approx 212$ MHz.}
    \label{fig:fock}
    \end{figure*}

A fundamental application of $H_{\mathrm{BS}}^{\mathrm{eff}}$ is the deterministic generation of high-photon-number Fock states. We demonstrate the synthesis of a $|4\rangle$ Fock state through a sequence of selective beam-splitter and two-mode squeezing operations, establishing the building blocks for more complex state engineering including binomial quantum error correction codes~\cite{Michael2016}.

\paragraph*{Protocol.}
The deterministic preparation proceeds via the following resonant pulse sequence:

\begin{enumerate}
    \item \textbf{Initial Pair Creation:} A TMS $\pi$-pulse resonant with $|0,0\rangle \leftrightarrow |1,1\rangle$ generates $\ket{1,1}$ from vacuum.
    \item \textbf{First Photon Transfer:} A BS $\pi$-pulse resonant with $|1,1\rangle \leftrightarrow |2,0\rangle$ transfers population to $\ket{2,0}$.
    \item \textbf{Second Pair Creation:} A TMS $\pi$-pulse resonant with $|2,0\rangle \leftrightarrow |3,1\rangle$ generates an additional photon pair.
    \item \textbf{Final Photon Transfer:} A BS $\pi$-pulse resonant with $|3,1\rangle \leftrightarrow |4,0\rangle$ completes the transfer to the target state $\ket{4,0}$.
    
\end{enumerate}

The performance of this protocol is quantified in Fig.~\ref{fig:fock}. Panel (a) shows the population dynamics during the four-pulse sequence, achieving coherent transfer to $\ket{4,0}$ with peak fidelity $\mathcal{F} = 0.997$. The Wigner function in panel (b) exhibits the characteristic signature of a pure Fock state, with concentric rings of alternating positivity and negativity. Panel (c) demonstrates robustness against experimental imperfections, maintaining high fidelity across practical parameter ranges.

\paragraph*{Remarks.}
This modular approach to Fock state synthesis provides a transparent and scalable alternative to numerically-optimized control pulses. The sequential application of selective interactions naturally extends to higher-photon-number states and forms the basis for synthesizing complex quantum codes and non-classical superpositions.

Effective Hamiltonians enable precise quantum state engineering across diverse settings. Unitary control via selective BS and TMS interactions facilitates the deterministic preparation of Fock states, superpositions, and entangled resources. Combined with dissipation, they autonomously stabilize nonclassical states against decoherence. These protocols underpin advanced applications in error correction, metrology, and simulation, establishing a foundation for scalable quantum technologies.

\section{Experimental Realizations and Fidelity Considerations}
\label{sec:experimental}

The selective control of individual quantum transitions is a cornerstone for high-fidelity quantum information processing. However, in multi-mode systems, this selectivity is challenged by spectral crowding and parasitic couplings. Our theoretical framework provides a direct solution by establishing the precise Hamiltonian conditions for engineering frequency-selective operations. In this section, we contextualize our findings by reviewing the experimental platforms where this framework is most applicable, focusing on how the requisite conditions for selectivity, primarily a hierarchy of energy scales, are realized in state-of-the-art systems, with particular emphasis on QED. 

\begin{figure}[b] 
    \centering
    \includegraphics[width=\columnwidth]{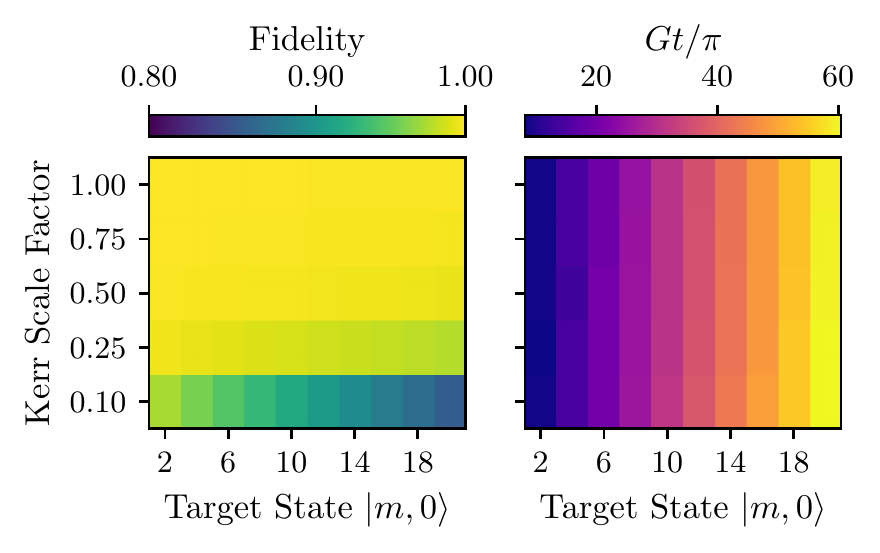}
 \caption{\justifying Impact of Kerr nonlinearity on the synthesis of Fock states $\ket{m,0}$.
    (a) Maximum state preparation fidelity as a function of target photon number $m$ and a relative Kerr scale factor. High fidelities ($>0.95$) are achieved for scale factors $\ge 0.50$.
    (b) Optimal preparation time in dimensionless units of $Gt/\pi$. The time scales linearly with the target number $m$ and is largely independent of the Kerr scale factor.
    Key parameters for the sequential protocol: $G/2\pi = J/2\pi = 20$ MHz, with base Kerr nonlinearities (scale=1.0) of $K_1/2\pi = 500$ MHz and $K_2/2\pi \approx 354$ MHz.}
    \label{fig:kerrn}
    \end{figure}

The influence of the Kerr nonlinearities on fidelity and operation time, summarized in Fig.~\ref{fig:kerrn}, directly illustrates the practical trade-offs encountered in experiments. Panel (a) shows that high fidelities ($\mathcal{F} > 0.98$) are maintained for relative Kerr scale factors above 0.5, confirming that moderate nonlinearities are sufficient for spectral selectivity. Panel (b) demonstrates that the optimal preparation time scales linearly with the target photon number and remains largely insensitive to the precise Kerr strength. These trends highlight the balance between nonlinearity, speed, and coherence that governs experimental feasibility, motivating the comparative discussion that follows.

\subsection{Platform Comparison and Engineering Trade-offs}

The two-mode Kerr Hamiltonian is a generic model realized in several physical platforms. The key parameter for selective control is the strength of the Kerr nonlinearity \(K\), which varies significantly across systems, leading to a direct trade-off between gate speed and coherence.

\begin{itemize}
    \item \textbf{Circuit QED:} Superconducting circuits, particularly transmon qubits, provide a native and high-fidelity realization of Kerr oscillators. The Kerr coefficient \(K\) is the transmon's intrinsic anharmonicity, with typical values of $K/2\pi \approx 200-400$ MHz \cite{Wei2023, Sears2012}. This large nonlinearity enables fast, selective gates and is the primary reason circuit QED is the leading platform for demonstrating the protocols in this work.

    \item \textbf{Trapped Ions:} In these systems, the motional modes are harmonic, and effective Kerr nonlinearities must be engineered using state-dependent optical forces. These induced nonlinearities are typically weaker, in the range of $K/2\pi \sim 1-100$ kHz \cite{Home2009}. While trapped ions offer exceptional coherence, the slower gate speeds imposed by weaker nonlinearities make high-speed selective control more challenging.

    \item \textbf{Optomechanical Systems:} Radiation pressure coupling in optomechanical systems can yield effective Kerr nonlinearities under strong driving, typically in the kHz to MHz range \cite{Rabl2011, Aspelmeyer2014}. The feasibility of selective control in these systems is highly dependent on achieving a sufficiently large $K/\kappa$ ratio.
\end{itemize}

Given its strong intrinsic nonlinearities and mature engineering toolkit, the circuit QED platform is the most direct and promising setting for implementing our framework. The following discussion will therefore focus on its implementation details.

\subsection{Implementation in Circuit QED}

In a circuit QED architecture, the two Kerr oscillators are realized as frequency-tunable transmon qubits. High-fidelity operation requires engineering the Hamiltonian parameters and verifying a specific hierarchy of energy scales, as summarized in Table~\ref{tab:params}.

\begin{table}[h!]
\centering
\caption{Typical experimental parameters for high-fidelity operations in circuit QED systems implementing the two-mode Kerr model.}
\begin{tabular}{@{} l c l @{}}
\toprule
\textbf{Parameter} & \textbf{Typical Value} & \textbf{Unit} \\
\midrule
Mode frequency, $\omega_j/2\pi$ & 5--10 & GHz \\
Kerr coefficient, $K_j/2\pi$ & 200--400 & MHz \\
Beam-splitter rate, $J/2\pi$ & 1--20 & MHz \\
Squeezing rate, $G/2\pi$ & 1--20 & MHz \\
Photon loss rate, $\kappa/2\pi$ & 1--100 & kHz \\
Dephasing rate, $\gamma_\phi/2\pi$ & 1--100 & kHz \\
\bottomrule
\end{tabular}
\label{tab:params}
\end{table}

\begin{table*}[t]
	\centering
	\caption{Key experimental demonstrations of two-mode control in circuit QED.}
	\label{tab:exp_summary}
	\begin{tabular}{ l l l l l }
		\toprule
		\textbf{Reference} & \textbf{Platform} & \textbf{Engineered Term} & \textbf{Demonstrated Protocol} & \textbf{Relevance to Selectivity} \\
		\midrule
		Wang \textit{et al.} (2016) \cite{Wang2016} & Circuit QED & Parametric Drive ($G$) & Two-mode Cat State Generation & \parbox[t]{3.2cm}{Required precise frequency control to selectively address pair-creation transitions while suppressing spurious interactions.} \\
		\hline
		Roushan \textit{et al.} (2017) \cite{Roushan2017} & Circuit QED & Beam-Splitter ($J$) & Quantum Simulation & \parbox[t]{3.2cm}{Demonstrated precise, site-selective control of photon hopping in a multi-transmon array, minimizing crosstalk.} \\
		\hline
		Ye \textit{et al.} (2020) \cite{Ye2020} & Circuit QED & Tunable $J$ and $G$ & Programmable Gate Set & \parbox[t]{3.2cm}{Directly engineered a universal gate set by dynamically activating and suppressing $J$ and $G$ interactions between two transmons.} \\
		\hline
		Sivak \textit{et al.} (2023) \cite{Sivak2023} & Circuit QED & Dissipative BS & Autonomous Stabilization & \parbox[t]{3.2cm}{Used a frequency-selective beam-splitter interaction to couple a data mode to a lossy ancilla, a key application of engineered dissipation.} \\
		\bottomrule
	\end{tabular}
\end{table*}

\subsubsection{Engineering the Hamiltonian}
\begin{itemize}
    \item \textbf{Kerr Nonlinearity ($K_1, K_2$):} The anharmonicity $K_j$ is an intrinsic property of the transmon. Fabricating transmons with deliberately different frequencies and anharmonicities ($K_1 \neq K_2$) is standard practice, which is crucial for breaking degeneracies and enabling selective addressing \cite{Wei2023}.

    \item \textbf{Linear and Nonlinear Couplings ($J, G$):} The beam-splitter interaction ($J$) is implemented via a capacitive coupling or a tunable coupler circuit \cite{Yan2018}. The two-mode squeezing interaction ($G$) is activated by parametrically driving a coupler element at a frequency near the sum of the two mode frequencies \cite{Puri2017}.
\end{itemize}

\subsubsection{Verifying the Hierarchy for Selectivity}
The theoretical framework demands a specific energy scale hierarchy for selective control:
\begin{enumerate}
    \item \textbf{Selectivity Condition ($|K_j| \gg J, G$):} Kerr nonlinearity must be the dominant energy scale to spectrally resolve transitions. This is confirmed via two-tone spectroscopy, which measures the anharmonicity \cite{Sokolov2024}. With $|K_j|/2\pi \sim 300$ MHz and $J, G \sim 20$ MHz, this condition is comfortably satisfied.

    \item \textbf{Coherence Condition ($J, G \gg \kappa_j, \gamma_{\phi,j}$):} Gate rates must far exceed decoherence rates to ensure coherent operations. Gate speeds of 10-50 MHz \cite{Ye2020} are achievable, while state-of-the-art decoherence rates are $\kappa, \gamma_\phi \lesssim 50$ kHz \cite{Martinis2015}.
\end{enumerate}

This results in the experimentally achievable hierarchy:
\begin{equation}
\begin{split}
    |K_j| \ (\sim 300\ \text{MHz}) & \gg \{J, G\} \ (\sim 20\ \text{MHz}) \\
    & \gg \{\kappa_j, \gamma_{\phi,j}\} \ (\sim 50\ \text{kHz}),
\end{split}
\end{equation}
which is the foundation for high-fidelity ($>99.9\%$) selective operations. Advanced pulse shaping techniques, such as Derivative Removal by Adiabatic Gate (DRAG)\cite{Motzoi2009}, are further employed to minimize spectral leakage and compensate for ac Stark shifts.

\subsection{Experimental Demonstrations of Selective Control}

The principles underlying our framework have been validated in several landmark circuit QED experiments that demonstrate control beyond single-mode effects, as summarized in Table~\ref{tab:exp_summary}.

For instance, the generation of entangled ``cat states'' across two microwave cavities relied on a parametrically activated $G$ term, requiring careful frequency management to ensure selectivity \cite{Wang2016}. Similarly, quantum simulation experiments with transmon arrays used the $J$ term to control photon hopping with minimal crosstalk \cite{Roushan2017}. Recent advances have culminated in programmable processors that use tunable couplers to implement a universal gate set based on the two-mode Kerr Hamiltonian \cite{Ye2020}, while other works have used selective dissipative couplings for bosonic qubit stabilization \cite{Sivak2023}.

In summary, circuit QED provides the most mature testbed for our framework, combining strong nonlinearities, highly developed control, and long coherence times. The experimental demonstrations to date validate the core principles of frequency-selective control in two-mode Kerr systems. The primary challenge moving forward is to extend these principles of selective control to multi-mode networks. In this context, the analytical framework and design principles established in this work provide a predictive roadmap for engineering the complex, crosstalk-resilient Hamiltonians required for next-generation, fault-tolerant bosonic processors.


\section{CONCLUSION}
We have presented a framework for achieving high-fidelity, selective quantum operations in a system of two coupled Kerr-nonlinear oscillators. The essential resource for selectivity is the anharmonicity provided by the Kerr effect, which renders individual Fock-state transitions spectrally addressable. High fidelity is achieved not merely by tuning the resonance condition of a target transition, but by ensuring that all parasitic transitions remain strongly off-resonant, with detunings much larger than their respective coupling strengths.

We identified a critical limitation for spectral selectivity arising from accidental degeneracies, which systematically occur when the ratio of Kerr coefficients $K_1/K_2$ is rational. This insight leads to a universal design principle of engineering a \textit{sufficiently complex rational ratio of Kerr nonlinearities} to eliminate unwanted resonances across the Hilbert space, thus allowing the preparation of large Fock states, for instance. Numerical simulations confirm that this approach enables deterministic control of Fock states up to $n=20$ with high fidelity ($>99\%$), demonstrating that the principal challenge at large photon numbers is the progressively longer interaction time required, rather than a breakdown of selectivity or coherence.

Finally, we discussed the relevance of these principles for realistic circuit-QED architectures, emphasizing the required hierarchy of energy scales and the need for compensating higher-order effects, such as AC Stark shifts, to reach fault-tolerant gate fidelities \cite{Fluhmann2019}. The framework introduced here provides a scalable toolkit for the deterministic engineering of complex entangled states and nonclassical resources. These capabilities are a prerequisite for advancing applications in quantum metrology \cite{Akamatsu2025}, quantum error correction \cite{Sivak2023}, and the autonomous protection of quantum states.

Crucially, the design principle of using incommensurate nonlinearities generalizes directly to multi-mode systems. In a network of $N$ coupled Kerr oscillators, accidental degeneracies arise when the nonlinear coefficients $K_{j}$ are rationally related. Ensuring these coefficients are rationally independent guarantees a sparse, non-degenerate transition spectrum, providing a clear architectural blueprint for scalable spectral addressability in large bosonic processors.

This work thus provides a clear roadmap for developing next-generation fault-tolerant quantum computing processors, while also inviting natural extensions to networks of $N$ coupled nonlinear oscillators, where selective multi-mode control and collective dissipation will be explored in future work.

\section*{Acknowledgments}
This work was supported by the Coordenação de Aperfeiçoamento de Pessoal de Nível Superior - Brasil (CAPES) - Finance Code 001, National Council for Scientific and Technological Development (CNPq), Grant 311612/2021-0, and by the S\~ao Paulo Research Foundation (FAPESP), Grant 2022/00209-6 and Grant 2022/10218-2.

\appendix
\refstepcounter{section}
\subsection*{Appendix A: High-Fidelity Control and Error Analysis via Magnus Expansion}
\label{appendix: A}

\begin{table*}[t]
\centering
\caption{Systematic error budget for a selective BS gate via Magnus expansion.}
\label{tab:magnus_errors}
\begin{tabular}{@{} l c c c l @{}}
\toprule
\textbf{Effect} & \textbf{Scaling} & \textbf{Magnus Order} & \textbf{Typical Magnitude} & \textbf{Compensation Method} \\
\midrule
Ideal Rabi oscillation & $\Omega$ & $\hat{\Omega}_1$ & $\sim$20 MHz & Primary gate operation \\
AC Stark shift & $|\Omega|^2/\Delta$ & $\hat{\Omega}_2$ & 100--500 kHz & Frequency calibration \\
Rabi freq. correction & $\Omega|\Omega|^2/\Delta^2$ & $\hat{\Omega}_3$ & 10--50 kHz & Amplitude calibration \\
Nonlinear Stark shift & $|\Omega|^4/\Delta^3$ & $\hat{\Omega}_4$ & 1--5 kHz & Higher-order correction \\
Coherent leakage & $\sim|\Omega|^3/\Delta^2$ & $\hat{\Omega}_3$ & $<0.01\%$ & Pulse shaping \\
\bottomrule
\end{tabular}
\end{table*}

To analyze and systematically suppress coherent errors that limit gate fidelity, we employ the Floquet-Magnus expansion~\cite{Blanes2009,Blanes2010}. This provides a rigorous method for constructing a time-independent effective Hamiltonian, $\hat{H}_{\text{eff}}$, that describes the stroboscopic evolution under periodic driving. The expansion generates the effective Hamiltonian as a series:
\begin{equation}
    \hat{H}_{\text{eff}} = \sum_{k=1}^\infty \hat{\Omega}_k,
\end{equation}
where each term $\hat{\Omega}_k$ involves $k$-nested commutators of the Hamiltonian, systematically incorporating higher-order effects.

To analyze the dynamics, we move into a frame rotating at the drive frequency, $\omega_{d}$, which is chosen to be resonant with a target BS transition, $\ket{n_0, m_0} \leftrightarrow \ket{n_0+1, m_0-1}$. In this frame, the Hamiltonian $\hat{H}'(t)$ naturally separates into a time-independent resonant component, $\hat{V}_{\text{res}}$, and a time-dependent part, $\hat{V}_{\text{off}}(t)$, which contains all off-resonant terms:
\begin{equation}
\hat{H}'(t) = \hat{V}_{\text{res}} +\hat{V}_{\text{off}}(t).
\end{equation}

The resonant term, $\hat{V}_{\text{res}}$, describes the desired coherent coupling within the targeted two-level subspace and is characterized by the Rabi frequency $\Omega^{(-)}_{n_0,m_0}$:
\begin{equation}
\hat{V}_{\text{res}} = \Omega^{(-)}_{n_0,m_0} \left( \ket{n_0+1, m_0-1} \bra{n_0, m_0} + \mathrm{H.c.} \right).
\end{equation}

In contrast, the off-resonant term $\hat{V}_{\text{off}}(t)$ includes all other transitions. These terms oscillate rapidly at their respective detunings $\delta_k$:
\begin{equation}
\hat{V}_{\text{off}}(t) = \sum_{k \neq \text{res}} \left( \hat{V}_k e^{i\delta_k t} + \hat{V}_k^\dagger e^{-i\delta_k t} \right).
\end{equation}

This separation is fundamental; under the rotating wave approximation (RWA), the terms in $\hat{V}_{\text{off}}(t)$ can be neglected if their detunings $\delta_k$ are sufficiently large, thereby isolating the desired beam-splitter dynamics.

\subsubsection{Systematic Error Analysis}

\textbf{First Order ($\hat{\Omega}_1$): Ideal Gate Operation.}
The first-order term is the time-averaged Hamiltonian, equivalent to the rotating wave approximation, which describes the ideal Rabi oscillations of the gate:
\begin{equation}
    \hat{\Omega}_1 = \hat{V}_{\text{res}}.
\end{equation}

\textbf{Second Order ($\hat{\Omega}_2$): AC Stark Shifts.}
This term introduces the dominant coherent error source: the AC Stark shift. It is a diagonal energy correction arising from virtual transitions to off-resonant states:
\begin{equation}
    \hat{\Omega}_2 = \hat{H}_{\text{Stark}}^{(2)} = \sum_{k \neq \text{res}} \frac{[\hat{V}_k, \hat{V}_k^\dagger]}{\Delta_k}.
\end{equation}
These shifts, scaling as $\mathcal{O}(|\Omega_k|^2/\Delta_k)$, must be compensated via frequency calibration to prevent gate phase errors.

\textbf{Third Order ($\hat{\Omega}_3$): Drive Renormalization.}
This term captures off-diagonal corrections, whose primary effects are a renormalization of the Rabi frequency (scaling as $\Omega_{\text{res}}|\Omega_k|^2/\Delta_k^2$) and the introduction of higher-order leakage channels.

\textbf{Fourth Order ($\hat{\Omega}_4$): Nonlinear Stark Shifts.}
For fidelities exceeding $99.99\%$, fourth-order corrections to the energy levels become essential, scaling as $\mathcal{O}(|\Omega_k|^4/\Delta_k^3)$.

\noindent
The hierarchy of coherent effects captured by the Floquet–Magnus expansion is summarized in Table~\ref{tab:magnus_errors}. Each term corresponds to a distinct physical mechanism that contributes to deviations from the ideal beam-splitter dynamics.

\subsubsection{Error Budget and Compensation Strategies}

The Magnus expansion provides a complete inventory of coherent errors and the strategies for their compensation.

\textbf{Coherent Phase Errors.} The primary phase error arises from the AC Stark shift, which must be compensated by adjusting the drive frequency. The total shift is the sum of all even-order diagonal corrections:
\begin{equation}
    \Delta E_{\text{Stark}}^{\text{total}} = \Delta E_{\text{Stark}}^{(2)} + \Delta E_{\text{Stark}}^{(4)} + \mathcal{O}(\Omega^6/\Delta^5),
\end{equation}
where $\Delta E_{\text{Stark}}^{(n)} = \bra{f} \hat{\Omega}_n \ket{f} - \bra{i} \hat{\Omega}_n \ket{i}$ for the gate's initial and final states.

\textbf{Population Leakage.} Leakage to off-resonant states is suppressed under the perturbative condition $|\Omega_k/\Delta_k| \ll 1$. To make this explicit, we define the detunings $\Delta_k$ from the target BS transition. For a parasitic BS transition, the detuning is determined by the Kerr coefficients:
\begin{equation}
    \Delta_k^{\text{BS}} = K_1(n' - n_0) - K_2(m' - m_0).
\end{equation}
For a parasitic TMS transition, the detuning is typically much larger and dominated by the mode frequencies:
\begin{equation}
    \Delta_k^{\text{TMS}} \approx 2\omega_2 + K_1(n' - n_0) + K_2(m' + m_0 - 1).
\end{equation}
The leading-order leakage probability can then be expressed in terms of fundamental parameters, explicitly showing the required hierarchy of energy scales:
\begin{multline}
    P_{\text{leakage}}^{(2)} = \sum_{(n',m') \neq (n_0,m_0)} \left| \frac{J\sqrt{(n'+1)m'}}{\Delta_k^{\text{BS}}} \right|^2 \\
    + \sum_{n',m'} \left| \frac{G\sqrt{(n'+1)(m'+1)}}{\Delta_k^{\text{TMS}}} \right|^2.
    \label{eq:leakage_explicit}
\end{multline}
Eq. ~\eqref{eq:leakage_explicit} makes the core requirement for high-fidelity control clear: the system must be in the strong dispersive regime where the Kerr coefficients are much larger than the coupling strengths, i.e., $|K_j| \gg J, G$. This ensures the detunings in the denominators are large, thus suppressing leakage.

\subsection*{Appendix B: Numerical Simulation Details}
\label{appendix: B}

Numerical computations throughout this work were carried out using the QuTiP (Quantum Toolbox in Python) library~\cite{qutip2}. In all simulations, including the population dynamics and robustness plots, the full Hamiltonian $\hat{H}_{\mathrm{Kerr}}$ from Eq.~\eqref{eq:H_full} was considered without resorting to the effective models, which were used only for analytical insight.

To ensure accurate results without expending unnecessary computational labor, we used a variable Fock basis size for each mode, $N_{\text{max}}$. The precision of the truncated basis was checked in each case by verifying the convergence of the results for all quantities studied. As a further check, we confirmed that the error in the bosonic commutator, $\langle [\hat{a}_j, \hat{a}_j^\dagger] \rangle - 1$, was smaller than $10^{-4}$ within the populated subspace.

To investigate the effects of environmental decoherence and thermal noise, we solved the Lindblad master equation~\cite{campaioli2024quantum} for the system density matrix $\hat{\rho}$:
\begin{equation}
    \dot{\hat{\rho}} = -i[\hat{H}_{\mathrm{Kerr}}, \hat{\rho}] + \sum_{j=1}^2 \mathcal{L}_j(\hat{\rho}).
\end{equation}
The dynamics are governed by the full Hamiltonian $\hat{H}_{\mathrm{Kerr}}$ and a sum of local dissipators $\mathcal{L}_j$ for each mode $j=1, 2$. Each dissipator describes photon loss (at rate $\kappa_j$) and coupling to a thermal bath (with mean occupation $n_{\text{th},j}$):
\begin{align}
    \mathcal{L}_j(\hat{\rho}) = &\ \kappa_j(n_{\text{th},j}+1) \mathcal{D}[\hat{a}_j](\hat{\rho}) \nonumber \\
    & + \kappa_j n_{\text{th},j} \mathcal{D}[\hat{a}_j^\dagger](\hat{\rho}),
\end{align}
where $\mathcal{D}[\hat{L}](\hat{\rho}) = \frac{1}{2}(2\hat{L}\hat{\rho}\hat{L}^\dagger - \hat{L}^\dagger\hat{L}\hat{\rho} - \hat{\rho}\hat{L}^\dagger\hat{L})$ is the standard Lindblad superoperator.

The use of this standard, local master equation is justified by the clear separation of energy scales in the system, as discussed in Sec.~\ref{sec:experimental}. The internal mode frequencies ($\omega_j$) and Kerr nonlinearities ($K_j$) are the largest energy scales, establishing the bare energy eigenstates (the Fock states). The inter-mode couplings ($J, G$) are significantly smaller, $J, G \ll \omega_j, |K_j|$. Most importantly, the dissipation rates are by far the smallest scale: $\kappa_j \ll J, G$. 

Because the dissipation is much weaker than all coherent dynamics, the standard Born-Markov approximations hold. Furthermore, since the system-environment coupling timescale ($\sim 1/\kappa_j$) is much slower than the inter-mode coupling timescale ($\sim 1/J, \sim 1/G$), it is not necessary to use a \textit{global or dressed-state} master equation~\cite{campaioli2024quantum}. The environment effectively couples to the individual, \textit{bare modes}, and the standard form with local dissipators $\hat{a}_j$ provides an accurate description of the system's open-quantum-system dynamics.

\bibliographystyle{unsrt}
\bibliography{main.bib}

\end{document}